\documentclass{article}

\usepackage{amsmath}
\usepackage{graphicx}
\usepackage{sidecap}
\usepackage{multirow}

    \usepackage[preprint]{neurips_2021}

\usepackage[utf8]{inputenc} 
\usepackage[T1]{fontenc}    
\usepackage{hyperref}       
\usepackage{url}            
\usepackage{booktabs}       
\usepackage{amsfonts}       
\usepackage{nicefrac}       
\usepackage{microtype}      
\usepackage{xcolor}         

\title{Object Based Attention Through Internal Gating}

\author{%
  Jordan Lei\\
  University of Pennsylvania\\
  \texttt{haochuan@seas.upenn.edu}\\
  \And 
  Ari Benjamin\\
  University of Pennsylvania\\
  \texttt{aarrii@seas.upenn.edu}\\
  \And 
  Konrad Paul Kording\\
  University of Pennsylvania\\
  \texttt{kording@seas.upenn.edu}\\
}

\begin{document}

\maketitle

\begin{abstract}
   Object-based attention is a key component of the visual system, relevant for perception, learning, and memory. Neurons tuned to features of attended objects tend to be more active than those associated with non-attended objects. There is a rich set of models of this phenomenon in computational neuroscience. However, there is currently a divide between models that successfully match physiological data but can only deal with extremely simple problems and models of attention used in computer vision. For example, attention in the brain is known to depend on top-down processing, whereas self-attention in deep learning does not. Here, we propose an artificial neural network model of object-based attention that captures the way in which attention is both top-down and recurrent. Our attention model works well both on simple test stimuli, such as those using images of handwritten digits, and on more complex stimuli, such as natural images drawn from the COCO dataset. We find that our model replicates a range of findings from neuroscience, including attention-invariant tuning, inhibition of return, and attention-mediated scaling of activity. Understanding object based attention is both computationally interesting and a key problem for computational neuroscience. 
\end{abstract}

\setcitestyle{numbers}

\section{Introduction}
Attention is the mechanism used by the brain to select a meaningful subset of a given stimulus \citep{attention} or features. This ability is a key component of visual perception and plays an important role in vision, learning, and memory \citep{attention, visualattention}. In the domain of vision, one well-studied type of attention is that of object-based attention, which refers to attention focused on an object apart from the rest of a visual scene.

A successful computational neuroscience model of object-based attention should work on nontrivial or even real world scenes. However, existing models of object-based attention in neuroscience have largely focused on simple recognition tasks and lack the expressivity to learn complex feature representations and dynamics. Because object-based attention fundamentally depends on both detecting objects and separating objects from their surroundings, attention developed in simple contexts cannot be naïvely applied to more complicated visual tasks.
Deep neural networks excel at working with complex scenes and may thus inform the search for models of object-based attention in the brain.

Here we enumerate the central characteristics of object-based attention that are neurobiologically relevant. We would like a model of object-based attention to share these salient facts with the brain.

\paragraph{Modulation of neural activation.}
Attention changes the way neurons respond to given stimuli. When a subject attends to an object in a region of the visual input, the neurons associated with that region will exhibit a stronger activation than when the subject is not attending to it \citep{attention, visualattention}. These changes typically fall on the order of 5 to 30\% changes in magnitude, with increasing attention modulation in deeper layers of the visual pathway \citep{attention}.

\paragraph{Attention invariant tuning.}
While object-based attention changes the magnitude of response, these changes do not alter the basic tuning characteristics of neurons \citep{mcadams1999effects, treue1999feature, attention}. Object-based attention can therefore be interpreted as a multiplicative scaling in the gain of neural responses \citep{mcadams1999effects, lee2010attentional, treue1999feature}, where neural tuning remains invariant to attention.

\paragraph{Internal gating.} One view of object-based attention involves internal gating through inhibition, where attention can be seen as `screening' or filtering out irrelevant signals for feature-driven or task-driven behavior \citep{desimone1995neural, olshausen1993neurobiological, olshausen1995multiscale}. It has been suggested that attention gating helps create a cleaner training signal for learning and credit assignment in the brain \citep{roelfsema2005attention}, and may play a central role in biologically plausible approximations of gradient descent \citep{qagrel}.

\paragraph{Hierarchical processing and learned feature representations.}
It has been suggested that the layered organization of cortical areas and the hierarchical functional organization of the ventral visual stream contribute to visual object processing and attention \citep{neocortex, attention, visualattention, felleman1991distributed, hubelandwiesel}. In the brain, these hierarchically organized cells often exhibit a wide diversity of tuning characteristics \citep{olshausen2005other}, which enable them to learn complex nonlinear feature representations. These feature representations capture salient characteristics of the input which inform attention in later layers \citep{attention, zelinsky2015and}.

\paragraph{Top-down neuromodulation of attention.}
In the visual pathway, information is propagated in the form of feedforward, feedback, and lateral connections \citep{lamme1998feedforward, macknik2009role}. It is commonly held that the role of feedback and lateral connections is to carry attentional modulation signals to earlier layers of the visual pathway, thus enabling top-down attention \citep{lamme1998feedforward, spratling2004feedback, attention, visualattention}. In this framework, feedforward connections are responsible for defining the receptive fields of earlier layers towards more complex representations in later layers. By contrast, horizontal and feedback connections, which have limited influence on the receptive fields, modulate tuning using attention by varying the relative magnitude of responses in relevant regions of the stimulus in earlier layers \citep{roelfsema1998object}.

\paragraph{Inhibition of return.} When humans attend to multiple objects in a visual scene, they typically do so sequentially, moving from one object to the next \citep{kochittireview, tsotsos1995modeling}. A wide range of experimental findings support the idea of inhibition of return (IOR), where recently attended regions of a visual input eventually become inhibited, allowing the subject to move from one object to the next \citep{iorklein2000inhibition, iorkwak1992consequences, iorposner1984components, kochittireview}. Without inhibition of return, a subject will remain fixated on the most salient object \citep{kochittireview}. Several computational models of this phenomenon have been proposed, most notably the routing control models of \citet{kochullman1984} and \citet{olshausen1993neurobiological, olshausen1995multiscale} which iteratively inhibit previously attended regions in multi-object attention tasks.

As we review in Section \ref{relatedwork}, the enumerated list of phenomena is not captured by methods in machine learning such as self-attention or general recurrence within the layers of a feedforward hierarchy, despite the common use of the word `attention'. 

Our work provides the following contributions: (1) A model for object-based attention which incorporates top-down, bottom-up, and lateral connections which learns visual representations of complex stimuli. Recurrent information flow, coupled with internal gating, drive top-down modulation of activity in earlier layers. (2) The use of this model in a task-driven learning context in which the model simultaneously learns to recognize and attend to objects in a visual scene. (3) Replication of neurobiologically relevant characteristics of object based attention, including attention invariant tuning, heterogeneous feature representations, internal gating and inhibition, top-down and bottom-up neuromodulation, and sequential allocation.

\section{Related work}
\label{relatedwork}

\subsection{Computational models of attention in the brain}
While many computational models of object-based attention have been proposed, most of them focus on extremely simple visual tasks, fail to learn complex feature representations over time, or fail to model inhibition of return. 

\citet{kochullman1984} represents early work proposing the use of pre-specified hierarchical saliency maps, a winner-take-all approach to selecting an object, and routing control to model inhibition of return (IOR). These principles were extended in the work of \citet{olshausen1993neurobiological, olshausen1995multiscale} to include scale-invariant saliency maps for dynamic IOR using routing control circuits. However, the pre-specified saliency maps severely constrain the ability of these models to represent complex stimuli and results in fairly homogeneous tuning characteristics, which limits the effectiveness of these models from both a computational and neurobiological standpoint.

Several variants of this archetype have been proposed. \citet{tsotsos1995modeling, itti1998model,sunfisher} propose hierarchical architectures with internal gating mechanisms for spatial and attentional selection, but do not learn to recognize objects. MORSEL \citep{morsel} combines attentional selection with object recognition, but is confined to simple recognition tasks of dots, bars, and circles. All of these models share similar shortcomings, namely, the pre-specified saliency maps and receptive fields constrain their expressivity, most of the models have rudimentary recognition modules (in some cases, none at all), and the IOR routing control dynamics are often imprecise.

\citet{deco2004neurodynamical, deco2005neurodynamics} take a slightly different approach, proposing a model of forward and reciprocal projections which are learned using local Hebbian learning rules. Despite the fact that the model can learn new connectivity, the model still relies on pre-specified Gabor filters, does not model inhibition of return, and is only applied to sparse arrangements of simple inputs.

\subsection{Object-based attention in deep learning}
Three concepts in deep learning have strong ties to the neuroscience of object-based attention: convnets (CNNs), recurrence, and encoder-decoder structures. The following models reproduce some, but not all, of the coarse features of object-based attention. These approaches can each be seen as informative steps towards a cohesive model of object-based attention (and inspire some of our research), but each falls short of providing a satisfying model of object-based attention in the brain.

The hierarchical organization of simple and complex cells in convnets (CNNs) is directly inspired by the work of \citet{hubelandwiesel} and enables CNNs to learn rich feature representations similar to those in the brain \citep{fukushima1988neocognitron, cadena2019deep,  khaligh2014deep, kriegeskorte2008representational, rajalingham2018large}. Attempts to understand the underlying computations constitute a type of `attention' by understanding what the networks are `looking at'. Methods such as Grad-CAM \citep{gradcam,guidedbackprop, zeiler2014visualizing} use fully-trained CNNs to create heatmaps of feature representations. These methods do not modulate neural activation, implement internal gating, describe IOR, or constitute top-down attention, all of which are key characteristics of object-based attention.

Similarly, Deep Attention Selective Networks \citep{dasnet} take fully-trained convolutional neural networks and learn to re-weight internal feature maps, modeling gating and multiplicative scaling. However, these post-hoc attention models fail to capture the way in which object-based attention can be used as a valuable training signal in the process of learning new feature representations. They also fail to model inhibition of return, which limits the model's effectiveness to single object-detection tasks. 

In segmentation models, objects are separated from the rest of a visual scene, similar to what is accomplished by the brain in object-based attention. Models such as RCNN \citep{rcnn}, Fast-RCNN \citep{fastrcnn}, and Faster-RCNN \citep{fasterrcnn} use region-proposal networks to generate bounding boxes around where an object could be. Mask-RCNN \citep{maskrcnn} runs a mask-proposal network to generate masks to isolate objects of interest in parallel with the bounding boxes. These models use segmentation as a training signal to better recognize and detect objects, but they still differ from object-based attention in the brain in important ways. First, there is limited evidence that the brain conducts region proposal, masking, and classification in distinct units like it does in those models. The use of bounding boxes for region proposals also has limited relevance to visual attention. In addition, these models lack feedback and lateral recurrent processing, which is thought to be critical for top-down attention and neuromodulation \citep{macknik2009role, lamme1998feedforward, attention, spratling2004feedback}.

There are several recurrent models of vision and visual attention. In CORnet-S \citep{kubilius2019brain}, the authors propose a recurrent convolutional model for brain-like object recognition. While the model serves as a powerful demonstration of the relevance of recurrence in modeling visual processing, it implements recognition without the use of an attention mechanism. Glimpse Net \citep{ba2014multiple} provides a model for the recognition of multiple objects in a visual scene, implicitly capturing routing dynamics through foveation. However, Glimpse Net is not a model of object-based attention because it fails to model neuromodulation of activation, internal gating, and top-down attention mechanisms.

The importance of feedback, feedforward, and lateral connections naturally lends itself to a discussion of a network architecture known as the U-Net \citep{ronneberger2015u}, which is characterized by an encoder-decoder architecture with an encoding "feedforward" pathway, decoding "feedback" pathway, and horizontal connections which give it a characteristic `U' shape. U-Nets have been successfully implemented as tools for segmentation, particularly for medical image segmentation \citep{ronneberger2015u, li2020anu, zhou2018unet++, oktay2018attention}, cell counting \citep{falk2019u}, and road detection \citep{yuan2019using}. One notable example is Focusnet \citep{focusnet}, a feedforward architecture that internal gating and attention mechanisms for medical image segmentation. However, FocusNet is a poor model of object based attention because it lacks recurrent information flow and does not do IOR, both of which are key features of object-based attention. While recurrent convolutional U-Nets have been applied to segmentation tasks \citep{r2unet, recurrentunet}, to our knowledge, no existing implementations of recurrent U-Nets utilize internal gating as a mechanism for information flow or have been developed as models of object-based attention in the brain.

\subsection{Relation to recurrent attention and self-attention}
In deep learning, the word `attention' has also been used to refer to the multiplicative re-weighting of intermediate outputs of encoder-decoder architectures, such as in  \citet{bahdanau2014neural} and \citet{xu2015show}. However, deep-learning notions of recurrent attention differ from neurobiological notions of attention. While these models capture the spirit of multiplicative scaling of activation, they typically do not employ internal gating and often fail to capture feedback projections from later to earlier convolutional layers. The use of a distinct `attention head' is also not supported by neuroscience.

In a similar vein, significant progress has been made in models of so-called `self-attention' in the field of deep learning, which removes the recurrent components of the network in favor of parallel processing of deep-learning-attention. In a review of neuroscience and deep learning approaches to attention, \citet{lindsay2020attention} notes that self-attention is actually less neurobiologically relevant relative to earlier notions of recurrent attention because it explicitly removes the recurrent information flow and lacks top-down mechanisms for attention. Despite the similarity in terminology, Bahdanau attention and self-attention in deep learning fail to capture key characteristics of object-based attention in the brain, particularly the top-down mechanisms of attention.

\section{Methods}
\label{methods}

\begin{figure}[t]
\centering
\includegraphics[width=0.9\textwidth]{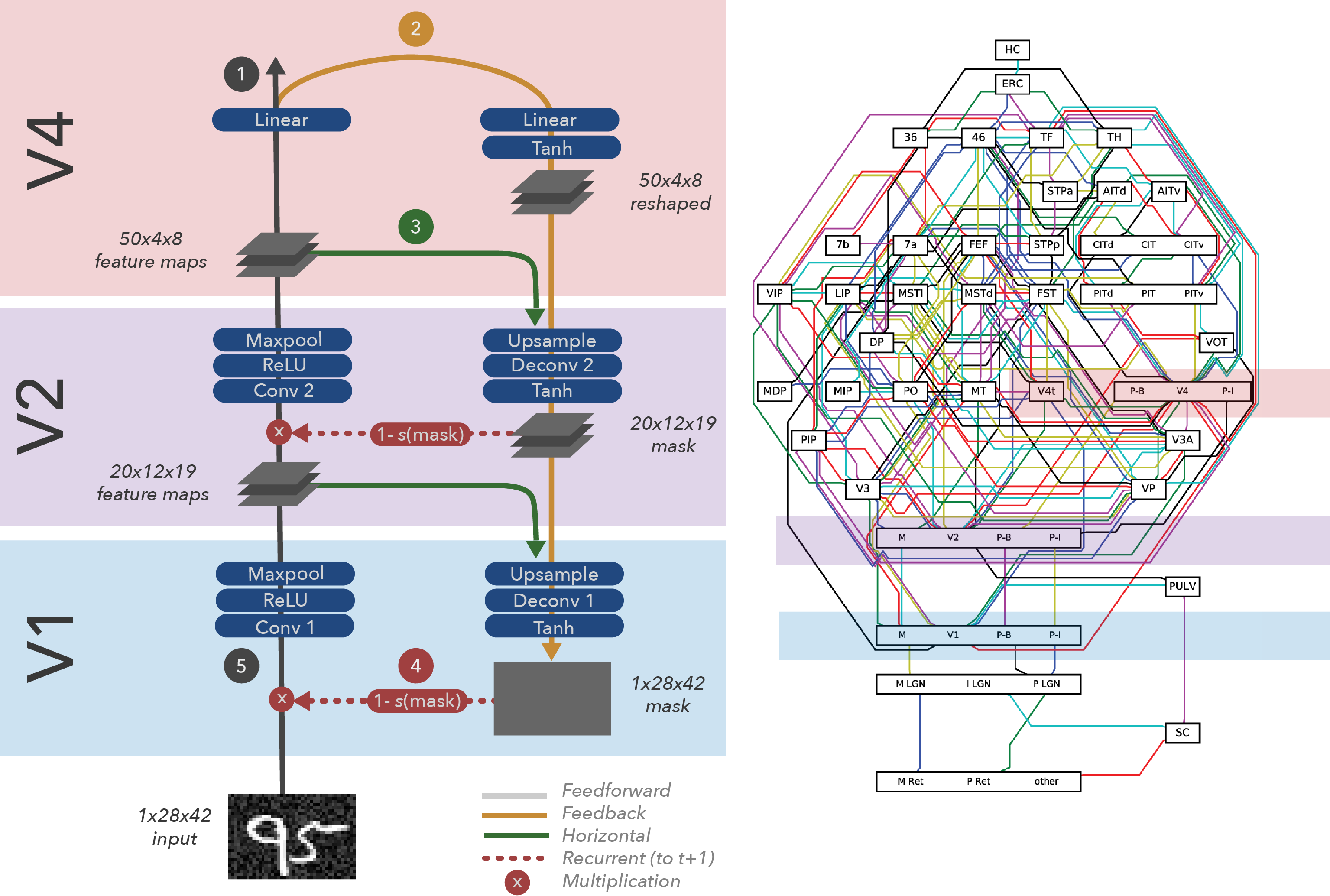}
  \caption{Object-based attention model. While real brains exhibit much more complex connectivity (right), our model (left) preserves feedforward, lateral, and feedback connectivity observed in the visual system. The connectivity diagram (right) is from \citet{felleman1991distributed} as depicted in \citet{jonas2017could}. Model properties: (1) The feedforward pathway generates feature maps and outputs a class prediction. (2) The feedback pathway generates attention masks. (3) Horizontal connections project from the feedforward pathway onto the feedback pathway at each layer. (4) Internal gating: attention masks are multiplicatively applied to the feature maps, where the parameter $s$ determines the strength of inhibition. (5) The final attention mask directly gates the input image, creating a pixel-space interpretable attended image.}
  \label{fig:modeldiagram}
\end{figure}

Based on existing neurobiological perspectives on visual attention, we propose a recurrent neural network, inspired by the U-Net architecture, which uses internal gating\footnote{Code for this project is available at \url{https://github.com/KordingLab/object-based-attention}}. The system is recurrent, combining bottom-up and top-down processing for each iteration. Bottom up activations are multiplicatively affected by top-down connections at each level (Figure \ref{fig:modeldiagram}). Architecturally, our model captures the feedforward, feedback, and lateral connections which are critical for object-based attention in the brain. Convolutional layers capture the hierarchical structure of the ventral visual pathway and are capable of learning complex feature representations.

To demonstrate that this model can both attend and recognize complex stimuli, we evaluate the model on tasks involving the selection of multiple objects. When the model selects an object, the object should become more salient relative to the remainder of the image. It should then classify that object and move onto other objects in the visual scene. 

\subsection{Model}

The goal of the feedforward pathway in our model is to create meaningful feature maps for recognition and attention, just like it seems to do in the brain. In our model, the feedforward pathway consists of the repeated application of convolutional kernels followed by nonlinearities and maxpooling layers. Each feedforward `block' generates a set of feature maps. The final feature map is flattened and passed to a linear decoder to generate a class prediction $\hat{y}$.

The goal of the feedback pathway is to take this encoding and create attention masks for all lower layers. It therefore consists of the repeated application of upsampling followed by deconvolution and nonlinear activations to construct an attention map which matches the shape of the feature maps created in the feedforward pathway. Horizontal connections project from the feedforward pathway to the feedback pathway, where feature maps of deeper layers are used to inform attention masks of earlier layers.

Recurrence is necessary for attention gating. Attention masks generated by the decoding pathway in iteration $t$ will be used to multiplicatively inhibit the feature maps in iteration $t+1$. A strength parameter $s$ serves as a scaling factor on the maximum strength of attention, $0 \leq s \leq 1$.

\subsection{Task}
We designed a minimal task framework for object-based attention. A single image which contains multiple objects is presented to the model. The goals of the model are to: (1) select objects using attention, and (2) classify the selected objects.

Each object in the sample therefore has a target `attended image' and target label corresponding with the two goals stated above. The `target attended image' for a given object looks identical to the original image, but all of the pixels outside the object are scaled by a factor of $(1-s)$ to represent the desired inhibitory effects of attention. In other words, the model is given a ground truth segmentation of the object for learning. In practice, the brain probably uses self-supervised learning for such tasks and could, for example, use common motion, feature similarity, contiguity, and other cues as an approximation to ground truth segmentations. The target label corresponds to the identity of that object. We thus have a simple task that may resemble the kinds of signals the brain can use to learn object-based attention.

The model will take the input image and process it for $T$ iterations. For our experiments we used $T = 5$. After $T$ iterations, it will output an attended version of the image (which we refer to as the "attended input" or "gated input", or $m_{1_T}$) and a class prediction $\hat{y_T}$. 

We infer which object the model selects by comparing the mean-squared-error of its attended input with the target attended images of each object in the sample, and picking the one which is minimal. To be consistent with our application of IOR (see Section \ref{ior}), we specify that once this object is chosen, it cannot be chosen again.

Our objective function can then be calculated to capture the two goals stated earlier. For the first, we use the mean-squared-error (MSE) calculation between the gated input from the model, $m_{1_T}$ and the target attended image of the selected object, $x$. For the second, we use a cross-entropy-loss calculation between the predicted label $\hat{y_T}$ and the ground-truth class label $y$. A parameter $\phi$ is used to adjust the relative weight between the two losses. Thus the loss can be summarized as follows: 
\begin{align}
    L = \phi MSE(m_{1_T}, x) + CrossEntropyLoss(\hat{y_T}, y)
\end{align}

\subsection{Inhibition of return}
\label{ior}
To handle multiple objects, our model implements inhibition of return (IOR), which has been observed in experimental contexts \citep{iorklein2000inhibition, iorkwak1992consequences, iorposner1984components}. The high-level idea of IOR is as follows: if we have attended to a particular object already, we should not attend to it again. 

The decoding pathway of the network contains the attention masks responsible for attending an object. These attention masks take on low values when an object is being attended to (minimally inhibited) and high values when they are not.

Once an object is selected, we use a threshold parameter $\theta$ to infer which features are being attended at each level of the network. The parts of each attention mask which fall below $\theta$ correspond to features that are being attended, and therefore should be inhibited in future iterations. We do this by setting the value of the attention mask (before scaling by $s$) to 1 for all future iterations.

\subsection{Dataset}
\label{dataset}
\subsubsection{Handwritten digits}
For our dataset, we use an augmented version of MNIST \citep{mnist}. Each sample consists of $n=2$ randomly selected digits, arranged side-by-side. White noise is added to the background of the image to force our model to learn to inhibit the background of the images as well. For each digit, the target image corresponds with the digit at full contrast with the rest of the image scaled by a factor of $(1-s)$, and the target label corresponds with the class of the digit, from zero to nine. For handwritten digits we use $s = 0.2$, reflecting the 5 to 30\% attentional scaling observed in experiments \citep{attention}.  

\subsubsection{Natural images}
We also apply our model to natural images drawn from the COCO dataset \citep{coco}, a large-scale object detection, segmentation, and captioning dataset. We filtered the dataset for images containing at least two non-overlapping objects of size greater than 10\% of the image area. We use the ground-truth segmentation masks from the COCO annotations to generate the target attended images by keeping the pixels within the segmentation mask at full contrast and scaling the pixels outside the image by a factor of $(1-s)$, as before. Given the increased difficulty of these images, we used $s = 0.3$ for a cleaner training signal and increased the number of layers of our model. An extended discussion of the modifications for the COCO model can be found in the Appendix.

For these images, the goal of the network was to sequentially attend to and classify the two largest objects in the visual scene. To mitigate some of the effects of class imbalance, we used the 12 coarse supercategories instead of the 80 object categories provided by COCO. These supercategories include ``accessory", ``animal", ``appliance", ``electronic", ``food", ``furniture", ``indoor", ``kitchen", ``outdoor", ``person", ``sports", and ``vehicle". Note that some of these labels correspond with well-delineated objects, such as ``person" and ``vehicle", while others, like ``indoor" and ``outdoor" do not. For this reason, we put less emphasis on the accuracy of the recognition aim of the model and more emphasis on its ability to attend to semantically meaningful objects.

\section{Results}
\subsection{Gated inhibition}
\label{gatedinhibition}
\begin{figure}
  \centering
  \includegraphics[width=0.8\textwidth]{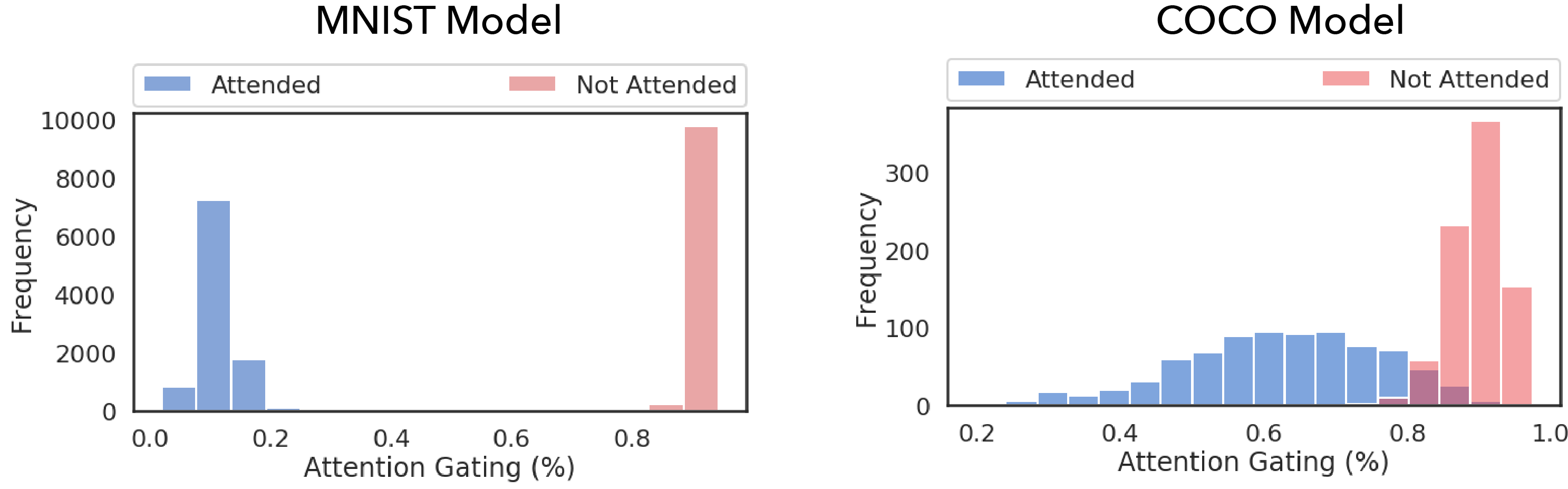}
  \caption{Gated inhibition histogram. In an ideal attention model, the attended regions should be fully uninhibited (0), and the non-attended areas should be fully inhibited (1). The histograms show the inhibition levels associated with attended objects (red) and non-attended surroundings (blue). The histograms are shown for the MNIST model (left) and the COCO model (right).}
  \label{fig:inhibition}
\end{figure}
\begin{figure}
  \centering
  \includegraphics[width=\textwidth]{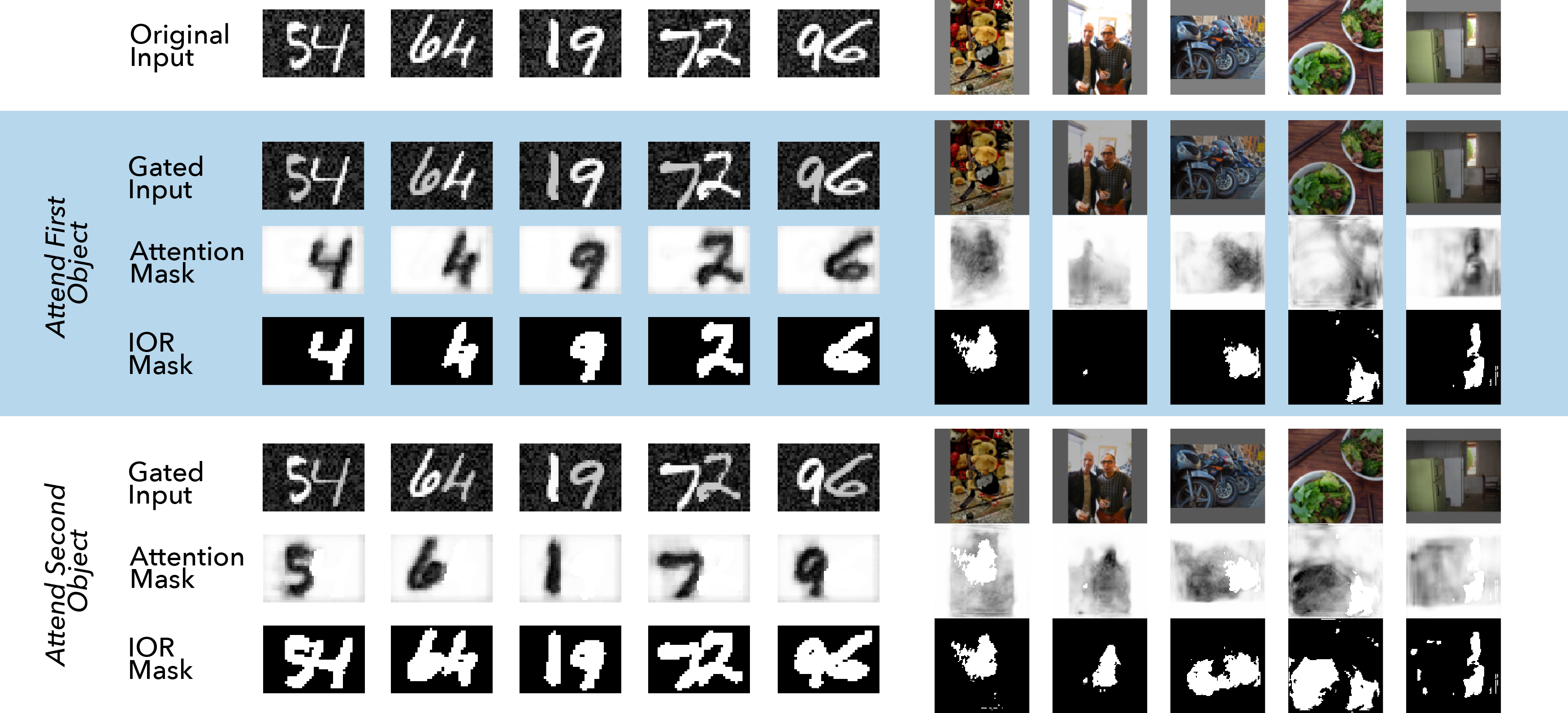}
  \caption{Visualization of attention gating. MNIST model (left) and COCO model (right) attending to objects in held-out images. The top row shows the original input. The gated input, attention mask, and IOR mask are shown for the selection of both objects. The gated input shows the effect of attention on the input image. The attention mask drives the gating process, where white regions correspond with inhibited areas and black regions correspond with uninhibited areas. The IOR mask specifies which regions have already been attended to, which will be inhibited in future iterations.}
  \label{fig:visualizations}
\end{figure}

If our model of object-based attention works as intended, it should strongly inhibit regions outside the object while minimally inhibiting the object of interest. We measure inhibition by taking the final attention mask of the feedback pathway of the network. For each image in the held-out set, we measured the average inhibition (before scaling by $s$) within the target object compared with the average inhibition outside the target object (Figure \ref{fig:inhibition}). Complete inhibition corresponds with a value of 1 and no inhibition corresponds with a value of 0. For the MNIST model, the attended objects receive an average inhibition of 0.11 (standard deviation $\sigma$ = 0.03), whereas the regions surrounding the objects receive an average inhibition of 0.99 ($\sigma$ = 0.01). For the COCO model, the attended objects receive an average inhibition of 0.62 ($\sigma = $ 0.14), whereas regions surrounding the objects receive an average inhibition of 0.90 ($\sigma$ = 0.04). For both models, a paired t-test showed significant ($p < 0.01$) difference in inhibition levels between attended and non-attended stimuli. The network attends near perfectly on MNIST and attends a decently well on COCO.

Object-based attention in our model has interesting dynamics (Figure \ref{fig:visualizations}). The model chooses one of the objects, attends it, and then moves on to the next one, driven by the inhibition of return (IOR). What we observe is that the network can sequentially attend to handwritten digits and objects in natural visual scenes, and that attention conforms to the shape of the object that is being attended. 


\subsection{Tuning curves}
\begin{figure}
  \centering
  \includegraphics[width=\textwidth]{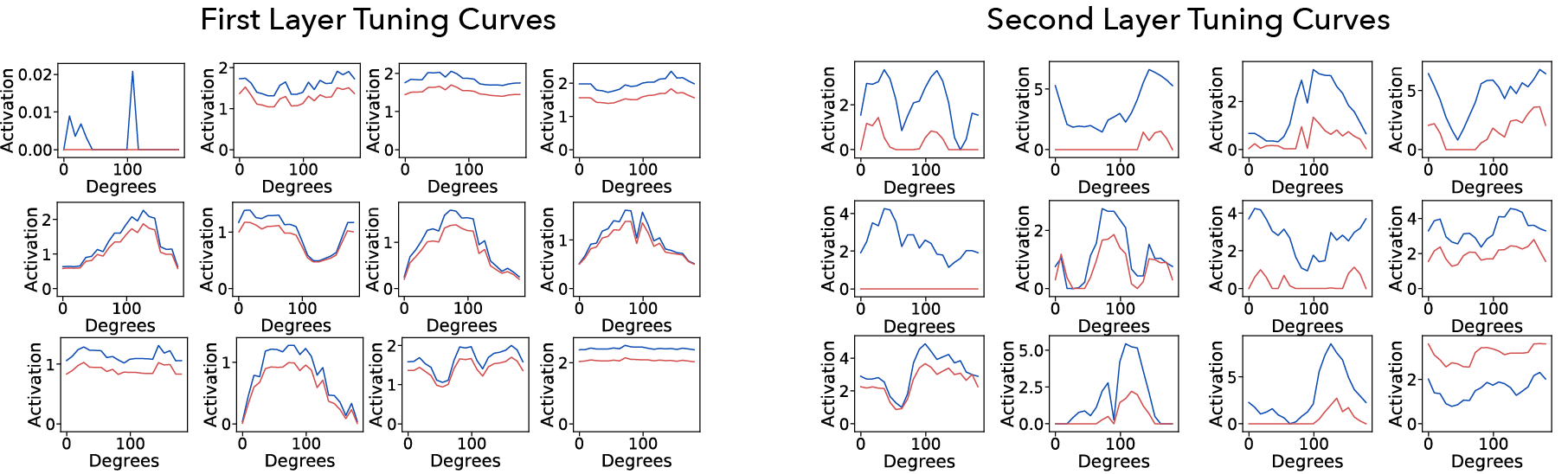}
  \caption{Tuning curves of activated neurons for the MNIST model. A representative sample of tuning curves for orientation from the MNIST model for attended (blue) and non-attended (red) stimuli. (Left) Tuning curves of neurons in the first convolutional layer demonstrate the multiplicative scaling of attention. (Right) Tuning curves of neurons in the second convolutional layer show increased modulation of attention. Rare cases exist (see bottom right) where this pattern is broken.}
  \label{fig:tuning}
\end{figure}

One of the key components of biological attention is the multiplicative scaling of activation in attended versus non-attended stimuli \citep{attention}. Tuning curves are often used to investigate this effect; using a simple stimulus, experimenters vary a specific feature, such as orientation, location, or size, and record neural responses. By plotting neural activation against the varying conditions of the stimulus, we can uncover what inputs drive neural activation \citep{hubelandwiesel}. In our model (Figure  \ref{fig:tuning}), just like in real neurons, attention multiplicatively scales tuning curves but does not change basic tuning characteristics \citep{attention, visualattention}.


An inspection of the tuning curves shows some similarities and differences from real neurons. Just like in real neurons, attention scales tuning while leaving the shape of tuning curves largely unaffected. Along with other models of tuning curve learning in the brain \citep{olshausen1996emergence}, they share with real neurons a level of specificity to the orientation of stimuli. Just like in real neurons \citep{attention}, the effect of attention appears to increase across the layers. And, just as in real neurons, there are some that buck the trend, being excited rather than inhibited by the gating process. However, tuning curves in real neurons tend to be smoother than those observed here. Neural activities in our attention system share important characteristics with neural activation in the brain.


\subsection{Accuracy}
\label{accuracy}
Our network should also be able to identify objects. In both datasets, we split the training data into training and validation sets using a $70-30$ split. Accuracy was measured on a per-object basis. The MNIST test set and the COCO validation set were held-out during training and used for final evaluation. On MNIST we achieved a training, validation, and test accuracy of $0.98$ on all three datasets.  To account for label imbalance in the COCO dataset, we evaluated accuracy on a per-label basis and took the mean across all labels in an equally weighted fashion. On COCO we achieved a training, validation, and testing accuracy of $0.37$, $0.27$, and $0.33$, respectively (baseline : $0.083$), using the supercategories.
As mentioned before, the comparatively low accuracy was due to our prioritization of attention over recognition. Our model is trained to both recognize and attend objects.


\section{Discussion}
Like many properties of visual processing in the brain, object-based attention in deep, hierarchical models is too complicated a phenomenon to program by hand. In this work we take a learning approach. We begin with an architecture that models characteristics which enable object-based attention in the brain, including feedforward, feedback, and lateral connections \citep{macknik2009role, spratling2004feedback}, internal gating \citep{kar2019evidence, spoerer2017recurrent, kubilius2019brain}, and inhibition of return \citep{kochullman1984, olshausen1993neurobiological, olshausen1995multiscale}. After training this model to sequentially attend and recognize objects, we observe several emergent characteristics of our model which are relevant to object-based attention in the brain. Specifically, we find that our model neurons multiplicatively attend to objects without changing the basic characteristics of tuning curves \citep{mcadams1999effects, lee2010attentional, treue1999feature}, neurons exhibit different levels of inhibition to attended versus non-attended stimuli \citep{attention, visualattention}, and deeper layers result in increased modulation of attention \citep{cook2002dynamics, attention}. We also observe the occasional exception to this general rule. These emergent properties of the model demonstrate its effectiveness in capturing the salient characteristics of object-based attention.




\section{Limitations and future work}
\label{limitations}
While our model captures a range of key characteristics of object-based attention in the brain, there are several aspects of the brain that our model does not yet capture.

Our model learns using backpropagation through time (BPTT). While backpropagation is a powerful tool for learning, there is no direct evidence that backpropagation is used as an updating mechanism in the brain \citep{lillicrap2020backpropagation}. However, biologically plausible approximations exist \citep{bengio2015towards} and models trained using backpropagation have been shown to learn brain-like representations \citep{cadena2019deep,  khaligh2014deep, kriegeskorte2008representational, rajalingham2018large, kubilius2019brain}.

While our model uses segmentations and labels supplied by supervision, the brain likely employs self-supervised learning \citep{deeplearning, anselmi2016unsupervised}. There are a range of ways this model could be adapted for self-supervised learning, for example using motion and contiguity cues for ground-truth attention.

There are also many aspects of neural computation we do not explicitly consider. In the brain, neurons compute nonlinear functions in their dendrites \citep{jones2021might}, and also have complex time-varying dynamics which are not modeled here. Attention mechanisms in the brain are also likely far more plastic.

In light of these limitations, we see several promising areas of future work. First, working with dynamic environments involving moving objects and temporary occlusion seems important. These effects have been well-studied in  neuroscience \citep{roelfsema1998object} and can be used to inform dynamic models of attention. Next, top-down attention can be studied as a mechanism to generate cleaner training signals in vision-related tasks, in line with existing work proposed by \citet{roelfsema2005attention}, potentially overcoming some of the texture-based biases that are known in traditional CNNs \citep{geirhos2018imagenet}. Finally, we envision extending this work to model reward-driven interactions involving attention, particularly in understanding the role of exploration and exploitation dynamics in visual search tasks \citep{ba2014multiple}.

\paragraph{Societal impacts.} It is also worth considering the potential societal risks of this work. Here we use the COCO dataset, which uses publicly available images, some of which depict people. While COCO does not release the personal information of the people depicted, faces and other visually identifiable features are present. Our model does not expose identifiable information beyond what is already publicly available through the dataset, which has been used widely as a benchmarking tool. More broadly, object-based attention has been proposed for use in facial recognition and surveillance technologies which violate privacy. Our model is not designed to make such fine-grained distinctions between individuals, and the authors do not condone the use of this model for surveillance and facial recognition purposes.

\section{Conclusion}
 Our model of object-based attention draws inspiration from neuroscience, including recurrent feedback, top-down and bottom-up neuromodulation of attention, and internal gating. We have demonstrated the model's ability to attend and classify multiple objects in its visual field. In addition, a wide variety of neurobiologically relevant characteristics of object-based attention emerge from our architecture, including tuning-invariant attention, inhibition of return, and increased modulation of attention in deeper layers. Our work takes a neurobiological approach to a computational model of task-driven visual object-based attention, motivating the continued cooperation between the fields of neuroscience and deep learning in the future. 

{
\bibliographystyle{plainnat}
\small
\bibliography{citations} 
}
\vfill
\pagebreak

\pagebreak
\appendix
\section{Appendix}
\subsection{Datasets}
\subsubsection{MNIST}
The original MNIST dataset \citep{mnist} consists of 60,000 training examples and 10,000 testing examples of high-contrast handwritten digits from 0 to 9. Each sample consists of an input, a $28 \times 28$ image containing a single handwritten digit, and a corresponding target label. The MNIST dataset is made available under the terms of the Creative Commons Attribution-Share Alike 3.0 license. Access to the dataset is available at  \url{http://yann.lecun.com/exdb/mnist/}

\begin{figure}[h!]
  \centering
  \includegraphics[width=\textwidth]{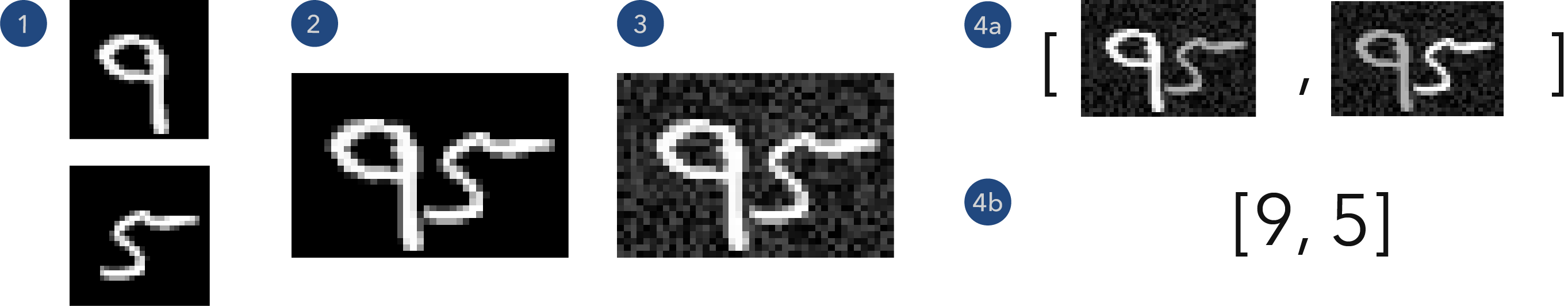}
  \caption{Augmentations to the MNIST dataset. (1) Two digits are randomly sampled from the MNIST dataset. (2) The digits are placed side-by-side with a 14-pixel offset. (3) White noise is added to the background. This image serves as the input to the model. (4a) Two target images correspond with the attended input for the left and right digit, respectively. (4b) Two target labels.}
  \label{fig:appendixmnist}
\end{figure}

We augmented the MNIST dataset to create the handwritten digit object recognition task (Figure \ref{fig:appendixmnist}). For each sample in our augmented dataset, we randomly select two digits from the MNIST dataset. In the case of train, val, and test subsets, we sample from only within the subset in question to ensure no validation or test samples are leaked to the training set. The two digits are arranged side-by-side with an offset of 14 pixels. White noise is added to the background (non-digit) portions of the image, drawn from $U[0, 0.3]$ for our experiments. The result is an image of size $28 \times 32$ which serves as the input. Each digit in the scene has a corresponding target label and a target image. The target image is the same as the input image except all the pixels not associated with the digit in question are scaled down by a factor of $(1-s)$, where $s$ is a strength parameter $0 \leq s \leq 1$. Note that $s = 1$ is effectively a segmentation task, where all regions outside the digit are completely masked. Thus, each sample consists of a single input image of size $28 \times 32$, two target labels corresponding to the classes of the digits present, and two target images also of size $28 \times 32$. In our experiments we set the value of $s = 0.2$, consistent with the 5 to 30\% range of attention modulation observed in the literature \citep{attention}. 

\subsubsection{COCO}
The COCO dataset is a large-scale object segmentation, detection, and captioning dataset \citep{coco}. Images from COCO are drawn from publicly available photos uploaded to Flickr. The COCO dataset is made available under the terms of the Creative Commons Attribution 4.0 License. Access to the dataset is available at  \url{https://cocodataset.org/#home}

We filtered the images, keeping only images that included at least two non-overlapping objects of size greater than $10\%$ of the image area. We then padded these images to conform to a square and resized them to $100 \times 100$ pixels. The filtered dataset contained 19033 training samples and 822 testing samples. To mitigate some of the effects of class imbalance, we used the 12 supercategory labels instead of the 80 object labels. While this helped alleviate some of the class imbalance, we still observed class imbalance in our dataset. We discuss strategies to address class imbalance in training and evaluating our model in Appendix Section \ref{appendixmodelparameters}.

We used the target segmentation masks from the COCO annotations to create versions of the image where the object of interest is a full contrast while the remainder of the image is scaled back by a factor of $(1-s)$, as before. Here we use $s = 0.3$. 

\subsection{Model}
\subsubsection{Model architecture}
\begin{table}[t]
 \caption{MNIST model architecture and parameters}
 \label{tbl:mnistmodel}
 \centering
 \begin{tabular}{lll} 
 \toprule 
 Layer Name &  Layer Type & Parameters \\ [0.5ex] 
 \midrule
 $conv_1$       & convolution     & in channels= 1, out channels= 20,\\ && kernel size = 4, stride = 1, padding = 0\\ 
 \midrule
 $maxpool_1$    & maxpool         & kernel size = 2\\
 \midrule
 $conv_2$       & convolution     & in channels= 20, out channels= 50,\\ && kernel size = 4, stride = 1, padding = 0\\ 
 \midrule
 $maxpool_2$    & maxpool         & kernel size = 2\\
 \midrule
 $linear_3$     & linear          & in size = 16000, out size = 10\\  
 \midrule
 $linear'_3$     & linear          & in size = 10, out size = 16000\\  
 \midrule
 $upsample_2$   & upsample          & in size = 4 x 8, out size = 9 x 16\\
 \midrule
 $deconv_2$     & convolution transpose  & in channels= 100, out channels= 20,\\ && kernel size = 4, stride = 1, padding = 0\\
 \midrule
 $upsample_1$   & upsample          & in size = 12 x 19, out size = 25 x 39\\ 
 \midrule
 $deconv_1$     & convolution transpose  & in channels= 40, out channels= 1,\\ && kernel size = 4, stride = 1, padding = 0\\
 \bottomrule
 \end{tabular}
\end{table}

\begin{table}[t]
 \caption{COCO model architecture and parameters}
 \label{tbl:cocomodel}
 \centering
 \begin{tabular}{lll} 
 \toprule 
 Layer Name &  Layer Type & Parameters \\ [0.5ex] 
 \midrule
 $conv_1$       & convolution     & in channels= 1, out channels= 30,\\ && kernel size = 4, stride = 1, padding = 0\\ 
 \midrule
 $maxpool_1$    & maxpool         & kernel size = 2\\
 \midrule
 $conv_2$       & convolution     & in channels= 30, out channels= 60,\\ && kernel size = 4, stride = 1, padding = 0\\ 
 \midrule
 $maxpool_2$    & maxpool         & kernel size = 2\\
 \midrule
 $conv_3$       & convolution     & in channels= 60, out channels= 100,\\ && kernel size = 4, stride = 1, padding = 0\\ 
 \midrule
 $maxpool_3$    & maxpool         & kernel size = 2\\
 \midrule
 $conv_4$       & convolution     & in channels= 100, out channels= 20,\\ && kernel size = 4, stride = 1, padding = 0\\ 
 \midrule
 $maxpool_4$    & maxpool         & kernel size = 2\\
 \midrule
 $linear_5$     & linear          & in size = 180, out size = 10\\  
 \midrule
 $linear'_5$     & linear          & in size = 10, out size = 180\\  
 \midrule
 $upsample_4$   & upsample          & in size = 3 x 3, out size = 6 x 6\\
 \midrule
 $deconv_4$     & convolution transpose  & in channels= 40, out channels= 100,\\ && kernel size = 4, stride = 1, padding = 0\\
 \midrule
 $upsample_3$   & upsample          & in size = 9 x 9, out size = 19 x 19\\
 \midrule
 $deconv_3$     & convolution transpose  & in channels= 200, out channels= 60,\\ && kernel size = 4, stride = 1, padding = 0\\
 \midrule
 $upsample_2$   & upsample          & in size = 22 x 22, out size = 45 x 45\\
 \midrule
 $deconv_2$     & convolution transpose  & in channels= 120, out channels= 30,\\ && kernel size = 4, stride = 1, padding = 0\\
 \midrule
 $upsample_1$   & upsample          & in size = 48 x 48, out size = 97 x 97\\ 
 \midrule
 $deconv_1$     & convolution transpose  & in channels= 60, out channels= 1,\\ && kernel size = 4, stride = 1, padding = 0\\
 \bottomrule
 \end{tabular}
\end{table}

Our model of object-based attention is inspired by the U-Net architecture \citep{ronneberger2015u}. The feedforward pathway consists of the repeated application of $4 \times 4$ convolutional kernels followed by ReLU nonlinearities and $2 \times 2$ maxpool layers. At the end of the encoding pathway, the feature maps are flattened and passed through a linear layer to generate a class prediction $\hat{y}$. On the feedback pathway, the latent representations are upsampled at each level and deconvolved with $4 \times 4$ deconvolutional kernels followed by tanh nonlinearities. At each level, horizontal connections project from the feedforward pathway to the feedback pathway, where the output of the maxpooling layer from the forward-projecting convolutional block is upsampled and concatenated to the input stream of the deconvolutional block.

\paragraph{MNIST model architecture.}
Table \ref{tbl:mnistmodel} describes the architecture of the object-based attention model trained on MNIST.

\paragraph{COCO model architecture.}
We changed our model in a few ways to account for the increased complexity of the images in the COCO dataset (Table \ref{tbl:cocomodel}). The main difference was the increased number of layers from 3 to 5. The upsampling interpolation was also changed from the default `nearest' to `bicubic', to allow for smoother upsampling for natural images. Instead of returning an output with 3 channels, the model returns an output with a single channel that is repeated 3 times along the channel dimension to match the input image RGB channels.

\subsubsection{Model training and hyperparameter selection}
\label{appendixmodelparameters}
The MNIST model was trained with a batch size of 64, $s = 0.2$, for 10 epochs. We used a grid search (Table \ref{tbl:gridsearch}) on the learning rate $lr$ and loss ratio $\phi$ parameters, for 5 trials each ($lr = 10^{-2}, 10^{-3}, 10^{-4}, 10^{-5}$, $\phi = 0, 10, 100, 1000, 10000$). As the model trained, we recorded the training accuracy as the mean accuracy over the past 100 batches. After the model trained, we evaluated the validation accuracy over the entire validation set. We selected the parameter combination which yielded the highest mean accuracy on the validation set over the 5 trials. We found that the best combination of parameters was $\phi = 1000$ and $lr = 10^{-3}$. The final model was selected from the 5 trials ($lr = 10^{-3}, \phi = 1000$) as the model which had the highest validation accuracy. The recorded test accuracy was the final model's performance on the entire held-out test set.

\begin{table}[t]
\caption{Grid search}
\label{tbl:gridsearch}
\centering
\begin{tabular}{lllllll}
                               & \textbf{}        & \multicolumn{5}{c}{Penalty}                                               \\
                               & \textbf{}        & {0.0} & {10.0} & {100.0} & {1000.0} & {10000.0} \\
                               \toprule 
\multirow{4}{*}{Learning Rate} & {0.00001} & 0.65     & 0.69     & 0.72      & 0.57        & 0.16         \\
\cmidrule(r){2-7}
                               & {0.00010} & 0.96     & 0.94     & 0.96      & 0.95        & 0.78        \\
                               \cmidrule(r){2-7}
                               & {0.00100} & 0.95     & 0.88     & 0.96      & \textbf{0.98  }      & 0.97        \\
                               \cmidrule(r){2-7}
                               & {0.01000} & 0.11     & 0.19     & 0.111     & 0.22        & 0.23        
\end{tabular}
\end{table}

The COCO model was trained with a batch size of 32, $s = 0.3$, for 100 epochs. Due to the computational and time constraints of the model, a grid search was infeasible. However, we tried a range of different parameter conditions and found that $\phi = 5000$ and $lr = 10^{-4}$
yielded favorable results on the validation set. Despite the fact that we used supercategories instead of the object categories provided, we still observed strong levels of class imbalance in our dataset. For example, approximately $38\%$ of the objects in the dataset corresponded with supercategory label `person'. To account for this label imbalance, we used a weighted cross-entropy in place of the normal cross-entropy measure in the loss function, where the weight of a given label was $1$ divided by the frequency of the label in the augmented COCO dataset. For a more balanced measure of performance, we evaluated the accuracy for each label and take the average per-label accuracy in an equally weighted fashion. The baseline performance using this equally-weighted accuracy metric is $0.083$. For train, validation, and test sets, we evaluated this metric after the model was completely trained.  

All models were trained on NVIDIA GeForce GTX TITAN X / NVIDIA GeForce RTX 2080 Ti GPUs. Training the MNIST attention model to convergence (around 10 epochs) takes under an hour. Training the COCO attention model to convergence (around 100 epochs) takes approximately 17 hours. 

\subsection{Loss function}
\label{appendixloss}
Recall that the loss function captures the two goals of the network: (1) select objects using attention, and (2) classify the selected objects:
\begin{align}
    L = \phi MSE(m_{1_T}, x) + CrossEntropyLoss(\hat{y_T}, y)
\end{align}
Where $\phi$ is a loss parameter that determines the relative importance of attending versus recognizing objects. As mentioned before, we use a weighted cross-entropy loss for the COCO model.

\begin{figure}
  \centering
  \includegraphics[width=0.8\textwidth]{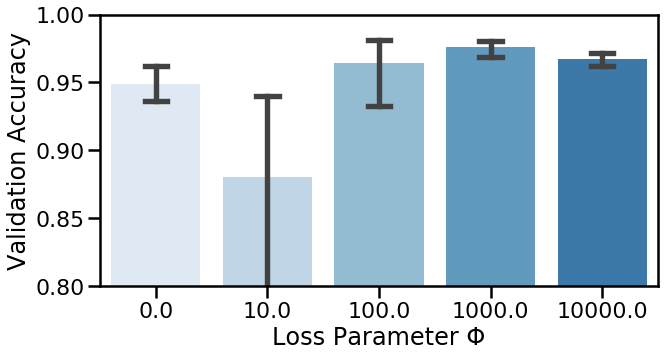}
  \caption{Impact of parameter $\phi$ on validation accuracy, $lr = 0.001$. Using the results of the grid search, we can visualize the impact of the gating loss penalty $\phi$ on the validation accuracy on the MNIST model. Error bars represent 95\% confidence intervals.}
  \label{fig:appendixgridsearch}
\end{figure}

Using the results of the grid search for the MNIST model, we can get a better understanding of how the choice of $\phi$ impacts validation accuracy (Figure \ref{fig:appendixgridsearch}). For values of $\phi$ that are too low, for example  $\phi = 10$, the penalty for the reconstruction loss is not high enough for the network to learn to use the internal gating mechanism to attend to digits properly. At the same time, this penalty term continues to prevent the network from directly optimizing for classification, which is why the validation accuracy of  $\phi=10$ tends to be lower than that of  $\phi = 0$. When  $\phi$ is too high, for example  $\phi = 10^5$, the gating loss penalizes the network to the point where it sacrifices classification accuracy in exchange for better internal gating and attention representations. $\phi = 1000$ represents a `sweet spot' where the two goals of attending and recognizing objects are well-aligned for the MNIST task. 

\subsection{Inhibition of return}
Inhibition of return is a key component of object-based attention \citep{kochullman1984, olshausen1993neurobiological, olshausen1995multiscale} and has been observed in experimental contexts \citep{iorklein2000inhibition, iorkwak1992consequences, iorposner1984components}. In our models, we use a threshold to determine which regions have been inhibited (not attended) or uninhibited (attended). If a region is uninhibited, that indicates that attention was given to this area and we should not attend to it again. We therefore use a threshold $\theta$, below which we say a given attention mask is uninhibited. For all future iterations we set the value of this attention mask (before scaling by $s$) to full inhibition (1), for all attention masks on the feedback pathway. In our experiments we used $\theta = 0.5$.

To be consistent with inhibition of return, we specify that an object that has been selected for evaluation (to calculate the loss) cannot be selected again. This creates a clean training signal and eliminates the risk of the model preferentially selecting for the `easier' of the two objects for recognition multiple times. 

One important implementation detail is that the attention masks of iteration $t$ are used to attend to an object in iteration $t+1$. For each object we run $T$ iterations, resulting in a gated input and class prediction. Thus, the attention masks used to inform the IOR process correspond with the model output of iteration $T-1$, the penultimate iteration. In our experiments we found $T = 5$ to be sufficient.

\subsection{Additional Figures}

\subsubsection{Visualizations}
\begin{figure}
  \centering
  \includegraphics[width=\textwidth]{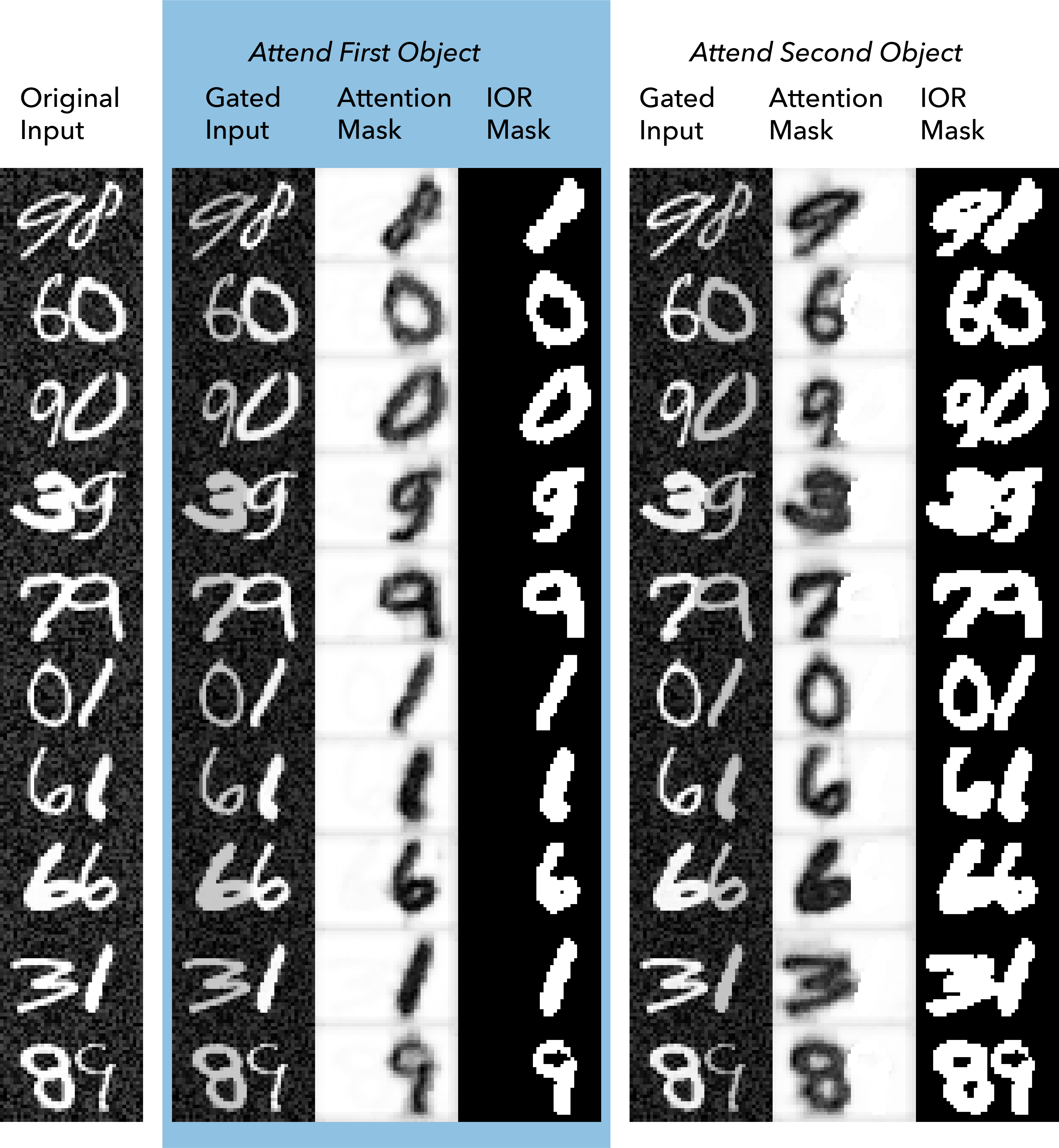}
  \caption{Additional visualizations of the MNIST model. Visualizations of the gated input, attention mask, and IOR mask for both objects in a sample of 10 held-out images.}
  \label{fig:appendixvismnist}
\end{figure}

\begin{figure}
  \centering
  \includegraphics[width=\textwidth]{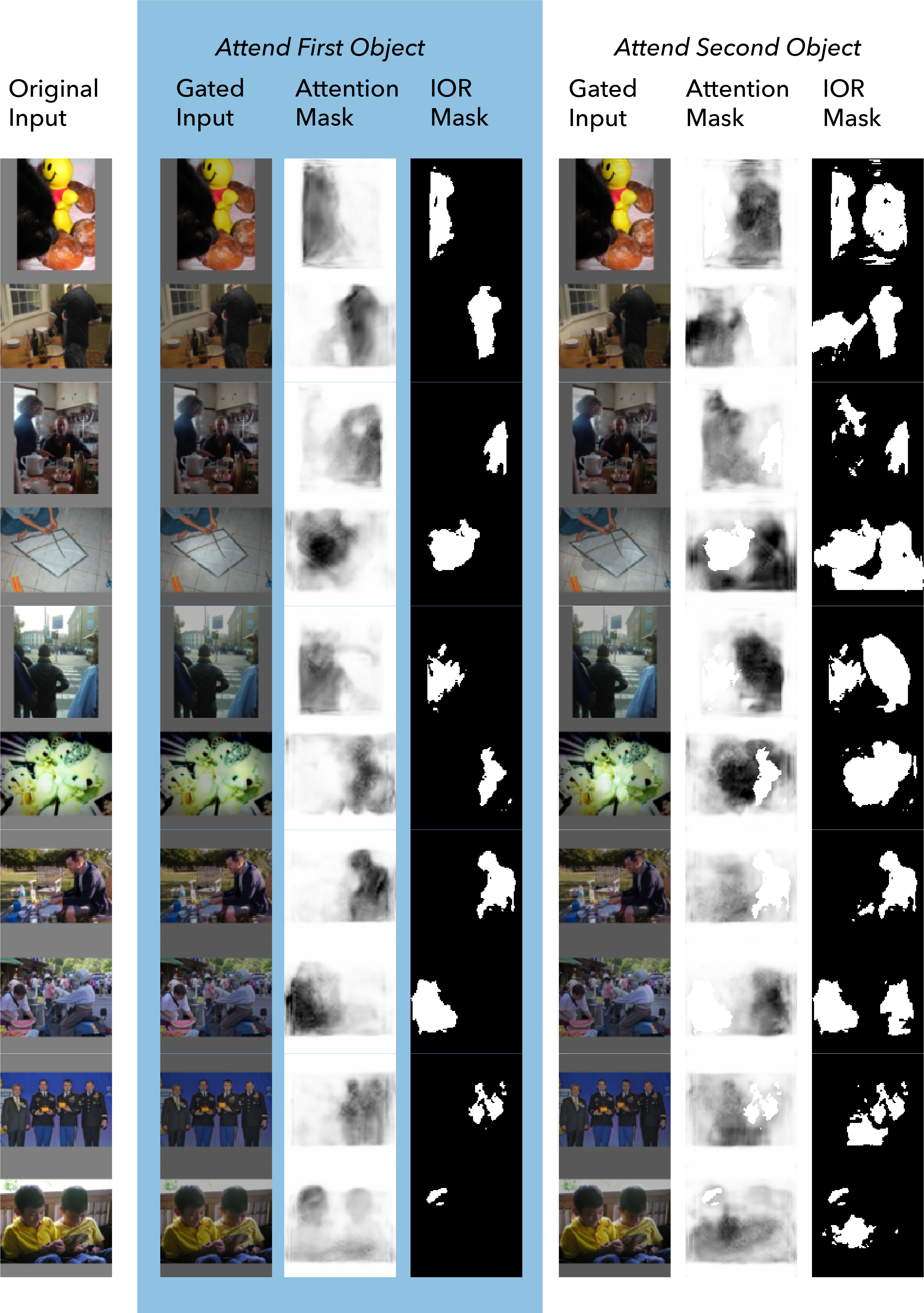}
  \caption{Additional visualizations of the COCO model. Visualizations of the gated input, attention mask, and IOR mask for multiple objects in a sample of 10 held-out natural images.}
  \label{fig:appendixviscoco}
\end{figure}

The attention model lends itself to intuitive visualizations of what the network is attending to. Additional visualizations for the MNIST and COCO model are shown in Figure \ref{fig:appendixvismnist} and \ref{fig:appendixviscoco}, respectively. The gated input represents the first-layer attention on the input image, demonstrating the scaling effects of attention in the first layer. The attention mask which multiplicatively scales the input can also be visualized, where inhibited (non-attended) regions appear white and uninhibited (attended) regions appear black. Finally, the attention masks are thresholded by $\theta$ to inform inhibition of return; these IOR masks enforce inhibition in these previously attended regions in future iterations.

These visualizations indicate that the attention masks conform to the shapes of the objects they attend to, a key feature of object-based attention. They also demonstrate the model's ability to select, attend, and sequentially inhibit objects. 

\subsubsection{Tuning curves}
Tuning curves are useful for investigating the effects of attention \citep{attention}. Tuning curves are generated by taking a simple stimulus and varying specific characteristics, for example, size, orientation, or location, and plotting the resulting neural activation under these varying conditions \citep{hubelandwiesel}. Attention modulates these tuning curves by scaling activation, but does not change basic tuning characteristics \citep{attention, visualattention}.

We modeled a stimulus as a horizontal bar of pixels. The horizontal bar was modeled using a 1-dimensional Gaussian repeated in the $x$ direction and normalized to the $[0, 1]$ value range. Using this horizontal bar, we varied the orientation, location (in both $x$ and $y$ directions), and bar width (which could be adjusted by the $\sigma$ value in the Gaussian). We placed this horizontal bar on the left half of the input image while adding a `distractor' input on the right half. This would allow the network to attend to the rotating bar during one phase and attend to the distractor during the next, allowing us to measure the activation of neurons during activated and non-activated phases.

\begin{figure}
  \centering
  \includegraphics[width=0.8\textwidth]{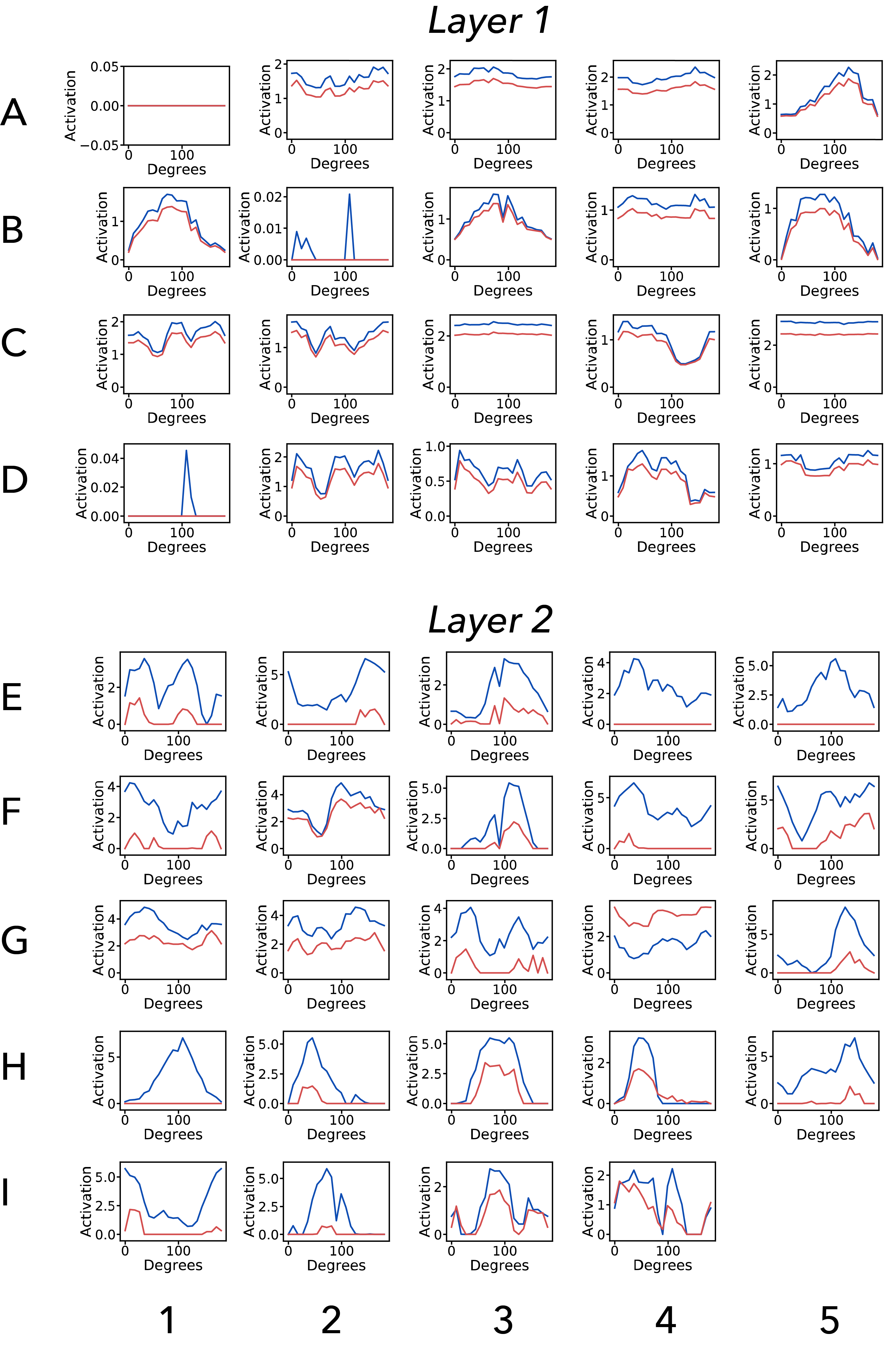}
  \caption{Tuning curves for activated neurons in MNIST model. Neurons that were not activated by any stimulus orientation (see A1) were removed. We observed 1 occurence of this in layer 1 and 26 occurences in layer 2. (Top) Layer 1 tuning curves exhibit characteristic multiplicative scaling of activation. (Bottom) Layer 2 tuning curves exhibit stronger attention modulation. Out of these neurons we found a single example which represents an exception to this rule (G4).}
  \label{fig:appendixtuning}
\end{figure}

The tuning curves were calculated for each orientation by taking the maximum activation of a neuron across all other conditions. Here, the output of a neuron corresponds to a single index of the feature map output of the convolutional layers in the model ($conv_1$, $conv_2$). We plot this relationship with the orientation (in degrees) on the x-axis and the activation on the y-axis. Tuning curves from layer 1 ($conv_1$) and layer 2 ($conv_2$) neurons are shown in Figure \ref{fig:appendixtuning}.

The tuning curves from our model share several characteristics with real neurons. Most notably, attention multiplicatively scales activation without changing the basic shape of the tuning curves. In deeper layers, we observe stronger modulation of attention, which is also observed in real neurons. We occasionally observe exceptions to this rule (see G4, Figure \ref{fig:appendixtuning}). This increased modulation at later layers may arise as a natural consequence of hierarchical gating, where subsequent layers compound the impact of inhibition. On the other hand, it is also possible that this increased modulation may arise from the feature selectivity of layers. In deeper layers of the network, simple features are composed into more specific and selective feature maps; this increased selectivity allows later layers to more easily differentiate between objects, accentuating the effect of attention in deeper layers of the network. In both of these explanations, the stronger modulation of attention is an emergent property of the architecture of the model after learning. 

\end{document}